\documentclass{jpsj-suppl}
\usepackage{txfonts} %Please comment out this line unless the txfonts package is availabe in your LaTeX system.
\usepackage{graphicx,amsmath,wrapfig,epsfig}
\title{Studies of exotic hadrons by high-energy exclusive reactions}
\author{
H. \textsc{Kawamura}$^a$, S. \textsc{Kumano}$^{a,b}$, 
and T. \textsc{Sekihara}$^a$
}
\inst{
$^a$KEK Theory Center, Institute of Particle and Nuclear Studies, KEK \\
\ \ 1-1, Ooho, Tsukuba, Ibaraki, 305-0801, Japan \\
$^b$J-PARC Branch, KEK Theory Center,
Institute of Particle and Nuclear Studies, KEK \\
\ \ and Theory Group, Particle and Nuclear Physics Division, 
J-PARC Center \\
\ \ 203-1, Shirakata, Tokai, Ibaraki, 319-1106, Japan
}
\recdate{July 15, 2013}

\abst{
We investigate the possibility of clarifying internal structure 
of exotic hadrons by high-energy exclusive reactions. 
In particular, the constituent-counting rule could be used
for determining the internal configuration in large-angle exclusive
scattering. As an example, we show the cross section 
$\pi^- + p \to K^0 + \Lambda (1405)$ in comparison
with the one for the ground-state $\Lambda$
production $\pi^- + p \to K^0 + \Lambda$. 
The counting rule indicates that the cross section scales as 
$s^{8} d\sigma /dt=$constant
if $\Lambda (1405)$ is an ordinary three-quark baryon, whereas
it is $s^{10} d\sigma /dt=$constant if $\Lambda (1405)$
is a five-quark baryon. Here, $s$ and $t$ are Mandelstam variables.
Such experiments could be possible at J-PARC,
LEP, JLab, CERN-COMPASS, and other high-energy facilities.
}

\kword{exotic hadron, exclusive reaction, perturbative QCD}

\begin{document}
\maketitle

$ \ \ \ $
\vspace{-1.2cm}
%%%%%%%%%%%%%%%%%%%%%%%%%%%%%%%%%%%%%%%%%%%%%%%%%%%%%%%%%%%%%%%%%%%%%%%%%%%%%%%%
\section{Introduction}
% \vspace{-0.2cm}

Hadrons are classified into two categories, mesons and baryons.
According to a naive quark model, the mesons and baryons have
the configurations of $q\bar q$ and $qqq$, respectively. 
Since the underlying fundamental theory of strong interactions (QCD)
does not prohibit different forms of configurations such as tetraquark
($qq\bar q \bar q$) and pentaquark ($qqqq\bar q$), exotic hadrons have 
been searched experimentally for a long time since 1960's.
%%%%%
It is fortunate that some exotic hadron candidates have 
been reported in the last several years particularly by 
the Belle and BaBar collaborations.
Nevertheless, it is rather difficult to find an undoubted evidence 
because similar theoretical results could be obtained for global
quantities such as spins, parities, masses, and decay widths even 
by conventional $q\bar q$ and $qqq$ models although there are some
indications, for example, that $f_0$(980) and $a_0$(980) could be
tetraquark or $K\bar K$ molecule 
and that $\Lambda \, (1405)$ could be a pentaquark or $\bar K N$ molecule
\cite{exotics}.

We have been investigating new approaches to the exotic-hadron studies by 
using high-energy hadron reaction processes 
\cite{kk-1589,kks-2013}, 
where quark and gluon degrees of freedom appear.
First, we consider the two-photon process $\gamma^* \gamma \to h \bar h$
for probing internal structure of the hadron $h$ such as
$f_0$(980) and $a_0$(980) \cite{kk-1589}. Exotic signatures appear in
generalized distribution amplitudes (GDAs) which
can be measured in the two-photon process.
The GDAs correspond to the GPDs (generalized parton distributions)
by the $s$-$t$ channel crossing.
The studies of the GDAs together with the GPDs should shed
light on a new aspect of exotic-hadron physics for future
developments in clarifying the existence of exotics
and their internal quark-gluon configurations \cite{kk-1589}.

Next, we propose to use hard exclusive production of an exotic hadron,
by taking $\Lambda \, (1405)$ as an example,
for finding its internal quark configuration \cite{kks-2013}. In particular, 
the cross section for the exclusive process $\pi^- + p \to K^0 + \Lambda (1405)$ 
is estimated at the scattering angle $\theta_{cm}=90^\circ$
in the center-of-mass frame by using exiting experimental data.
We suggest that the internal quark configuration of $\Lambda (1405)$
should be determined by the asymptotic scaling behavior of 
the cross section in comparison with the ground-state $\Lambda$ production.
Such measurements will be possible, for example, by using
the high-momentum beamline of J-PARC and at LEPS, JLab, 
and CERN-COMPASS.
Since the first project of two-photon process has not been 
completed yet, we discuss the second one on the constituent
counting rule \cite{kks-2013} in the following.

%%%%%%%%%%%%%%%%%%%%%%%%%%%%%%%%%%%%%%%%%%%%%%%%%%%%%%%%%%%%%%%%%%%%%%%%%%%%%%%%
\section{Constituent-counting rule for hard exclusive processes}

\begin{figure}[b]
\begin{minipage}{0.48\textwidth}
   \begin{center}
     \includegraphics[width=3.5cm]{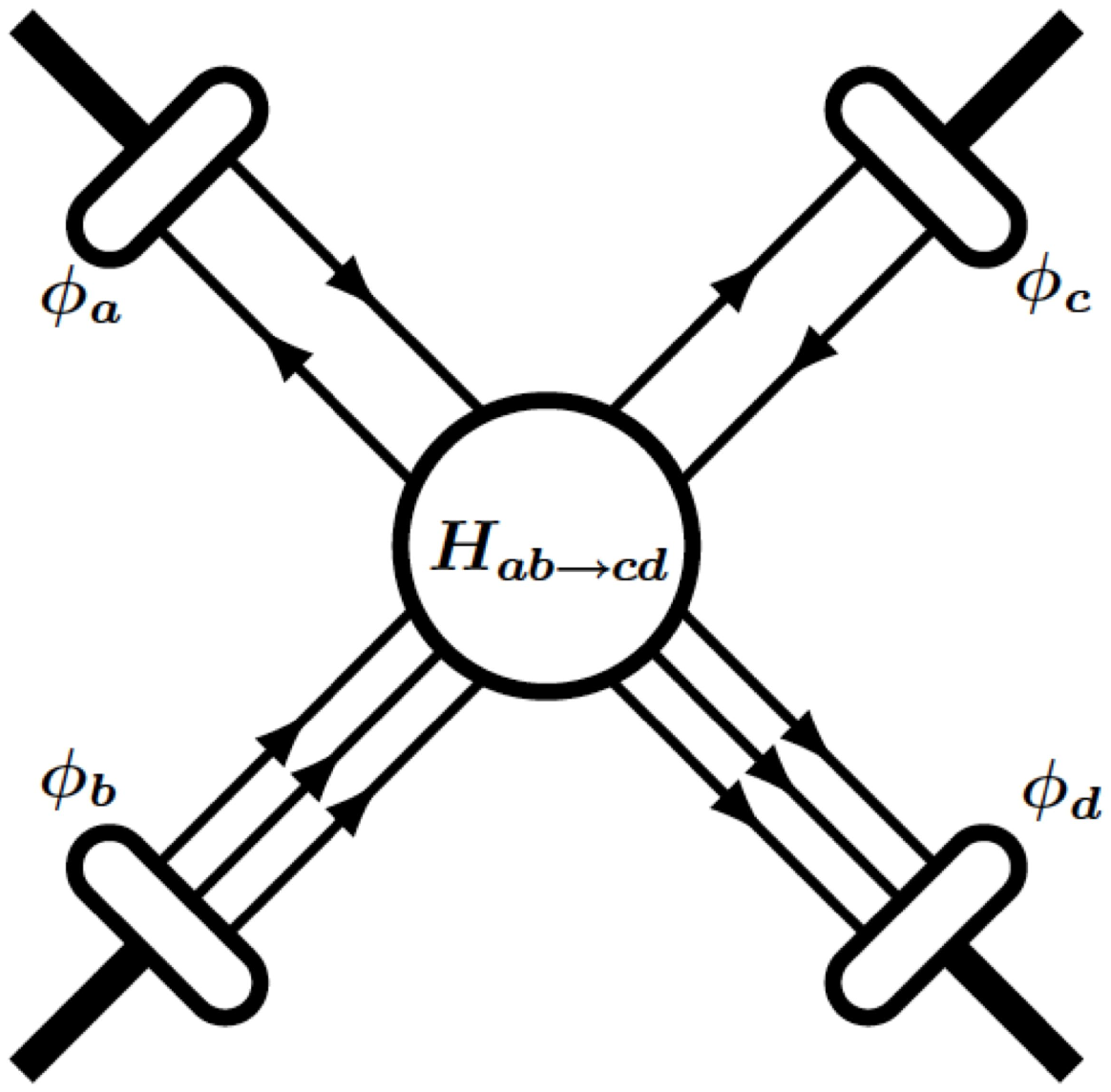}
   \end{center}
\vspace{-0.1cm}
\caption{Exclusive processes $a+b \to c+d$.}
\label{fig:exclusive-ab-cd}
\end{minipage}
%%%%%
\hspace{0.5cm}
\begin{minipage}{0.48\textwidth}
   \begin{center}
     \includegraphics[width=4.5cm]{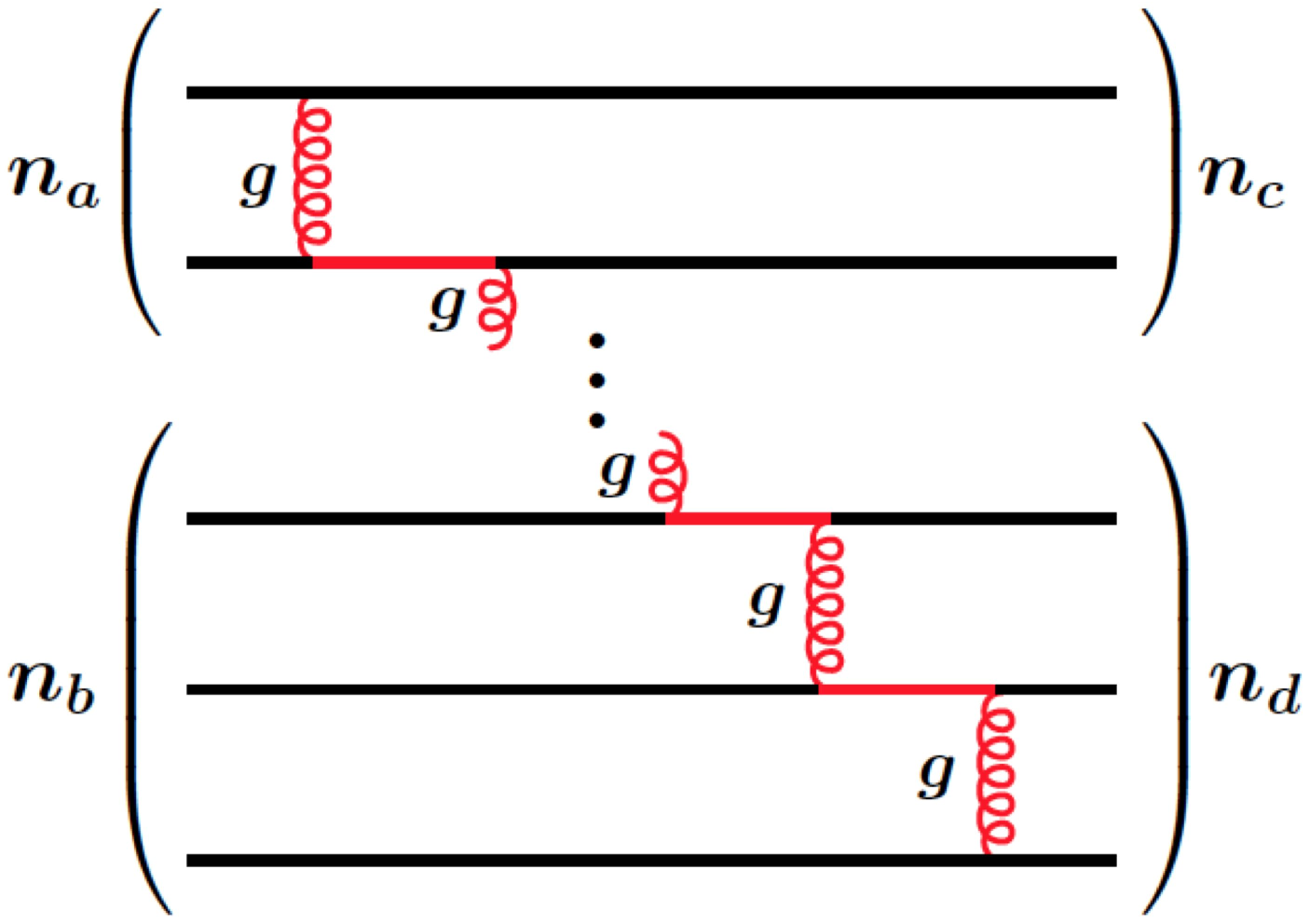}
   \end{center}
\vspace{-0.0cm}
 \caption{Hard gluon exchange process.}
% \caption{Hard gluon exchange process for an exclusive hadron-hadron
% reaction.}
\label{fig:hard-glun-exchange}
\end{minipage} 
\end{figure}

We consider a large-angle exclusive scattering $a+b \to c+d$.
Its cross section is given by
\begin{align}
\frac{d\sigma_{ab \to cd}}{dt}
& \simeq \frac{1}{16 \pi s^2}
\overline{\sum_{pol}} \, | M_{ab \to cd} |^2 ,
\label{eqn:two-body-cross}
\end{align}
where $s$ and $t$ are Mandelstam variables defined by
the momenta $p_i \ (i=a,\,b,\,c,\,d)$ as
$s = (p_a + p_b)^2$ and $t = (p_a - p_c)^2$,
and the matrix element is expressed as
\cite{matrix-abcd}
\begin{align}
M_{ab \to cd} = & \int [dx_a] \, [dx_b] \, [dx_c] \, [dx_d]  \,
    \phi_c ([x_c]) \, \phi_d ([x_d]) 
\nonumber \\
& \ \ \ 
\times 
H_{ab \to cd} ([x_a],[x_b],[x_c],[x_d],Q^2) \, 
      \phi_a ([x_a]) \, \phi_b ([x_b]) .
\label{eqn:mab-cd}
\end{align}
Here, $H_{ab \to cd}$ is the partonic scattering amplitude,
$\phi_i$ is the light-cone distribution amplitude of 
the hadron $i$ $(i=a,\,b,\,c,\,d)$ as
illustrated in Fig. \ref{fig:exclusive-ab-cd}, and
 $[x]$ indicates a set of the light-cone momentum 
fractions of partons in a hadron: $x_i=p_i^+/p^+$ where $p_i$ and
$p$ are $i$-th parton and hadron momenta, respectively.

%%%%%%%%%%%%%%%%%%%%%%%%%%%%% figure %%%%%%%%%%%%%%%%%%%%%%%%%%%%%
%%%%%%%%%%%%%%%%%%%%%%%%%%%%% figure %%%%%%%%%%%%%%%%%%%%%%%%%%%%%

For a particle $h$ with $N_h$ constituents, the state vector
could be written as 
$ | \, h \, \rangle = \sqrt{N_h} \, | \, n_h \, \rangle $.
The state vector of a hadron is normalized as
$\langle \, h(p') \, | \, h(p) \, \rangle
  =2p^0 (2\pi)^3 \delta^{(3)} (\vec p-\vec p\, ')$,
so that its mass dimension is $[1/M]$.
If the state vector of each constituent has the mass dimension $[1/M]$,
the normalization factor has the mass dimension 
$[\sqrt{N_h}] = [M^{n_h-1}]$.
The normalization factors are expressed by soft constants of the order 
of a hadron mass, so that other mass-dimension factors come from
hard momenta in the reaction.
The variable $s$ could be chosen as the only hard scale 
in the large-angle exclusive reaction, then the cross section becomes 
\cite{counting}
\begin{align}
\frac{d\sigma_{ab \to cd}}{dt} = \frac{1}{s^{\, n-2}} \, f_{ab \to cd}(t/s),
\label{eqn:cross-counting}
\end{align}
where $f(t/s)$ is a scattering-angle dependent function, and
the factor $n$ is defined by $n = n_a+n_b+n_c+n_d$.
Because the cross section scales as $1/s^{\, n-2}$ with the number 
of constituents, this scaling behavior is called the 
constituent-counting rule. 

This counting rule was investigated in perturbative QCD \cite{matrix-abcd}.
Due to the nature of the large-angle-exclusive reaction, quarks should 
share large momenta so that they should stick together to form
a hadron by exchanging hard gluons as shown in 
Fig. \ref{fig:hard-glun-exchange}.
Assigning hard momentum factors for the internal quark and 
gluon propagators and considering the hard factors for
external quarks, we could explain the constituent-counting rule
in perturbative QCD although there are some complications
from disconnected diagrams \cite{kks-2013}.

%%%%%%%%%%%%%%%%%%%%%%%%%%%%%%%%%%%%%%%%%%%%%%%%%%%%%%%%%%%%%%%%%%%%%%%%%%%%%%%%
\section{Exclusive production of exotic hadron
for probing its internal structure}

%%%%%%%%%%%%%%%%%%%%%%%%%%%%%%%%%%%%%%%%%%%%%%%%%%%%%%%%%%%%%%%
\begin{wrapfigure}{r}{0.42\textwidth}
   \vspace{-0.8cm}
   \begin{center}
     \includegraphics[width=6.0cm]{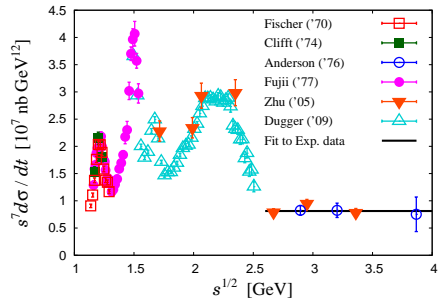}
   \end{center}
\vspace{-0.3cm}
\caption{Cross section of $\gamma + p \to \pi^+ + n$.}
% at $\theta_{c.m.}=90^\circ$.}
\label{fig:gamma-p}
\vspace{-0.5cm}
\end{wrapfigure}
%%%%%%%%%%%%%%%%%%%%%%%%%%%%%%%%%%%%%%%%%%%%%%%%%%%%%%%%%%%%%%%

The constituent-counting rule has been investigated experimentally
at BNL and JLab, and their data support the scaling predicted by
the counting rule \cite{exp-scaling}. As an example, the cross section 
$s^7 d\sigma/dt$ is shown in Fig. \ref{fig:gamma-p}
at $\theta_{c.m.}=90^\circ$ for the process $\gamma+p \to \pi^+ +n$,
where the total number of constituents is nine,
as a function of the center-of-mass
energy $\sqrt{s}$.
At low energies, resonance-like bumps appear, whereas
the cross section multiplied by $s^7$ is constant at high energies, 
which agrees with the scaling of the counting rule.
It is also interesting to find that the transition from
hadron degrees of freedom to quark ones seems to be 
apparent at $\sqrt{s}=2.5$ GeV.

We use the idea of constituent-counting rule for probing internal structure
of exotic hadron candidates, especially $\Lambda \, (1405)$.
The $\Lambda (1405)$ is a baryon resonance with 
isospin 0, spin-parity $(1/2)^-$, strangeness $-1$, 
mass 1405.1 MeV, and width 50 MeV.
The naive quark model 
treats the $\Lambda \, (1405)$ as an $uds$ three-quark system, 
but the quark model cannot explain that the $\Lambda \, (1405)$ mass 
is much ligher than that of the lightest nucleon resonance 
with $(1/2)^-$, $N(1535)$.  Therefore, instead of an $uds$ quark system,
it is thought as an exotic hadron such as a $\bar K N$ molecule
\cite{exotics}.
% Using the counting rule, we could clarify
% whether $\Lambda (1405)$ is an ordinary three-quark baryon or
% an exotic five-quark baryon including the $\bar K N$ molecule.

%%%%%%%%%%%%%%%%%%%%%%%%%%%%% figure %%%%%%%%%%%%%%%%%%%%%%%%%%%%%
\begin{figure}[b]
\begin{minipage}{0.48\textwidth}
   \begin{center}
     \vspace{-0.0cm}
     \includegraphics[width=6.0cm]{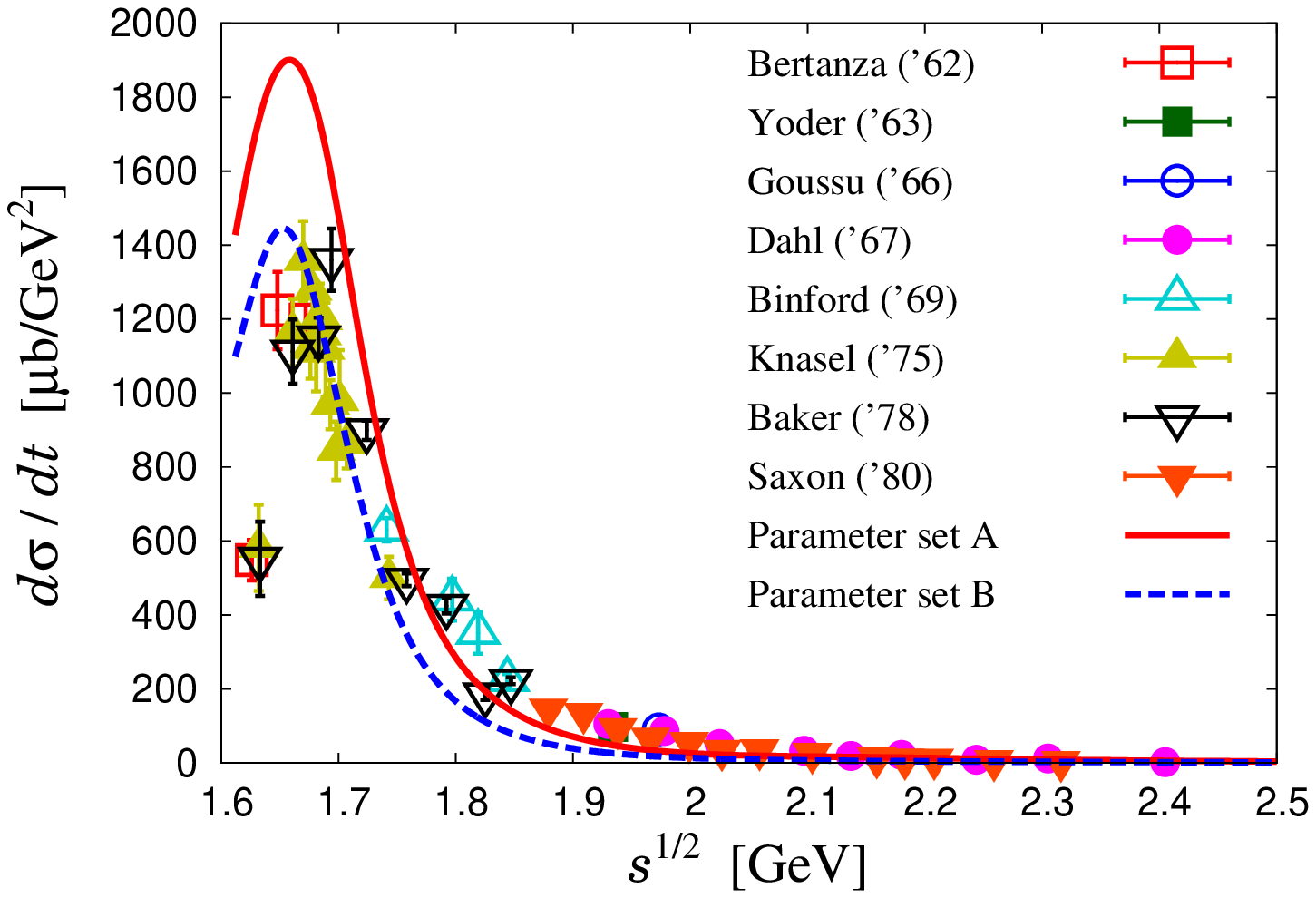}
   \end{center}
\vspace{-0.3cm}
\caption{Cross section of $\pi^- + p \to K^0 + \Lambda$ \cite{kks-2013}.}
\label{fig:lambda-pro-0}
\end{minipage}
%%%%%
\hspace{0.5cm}
\begin{minipage}{0.48\textwidth}
   \begin{center}
     \vspace{-0.0cm}
     \includegraphics[width=6.0cm]{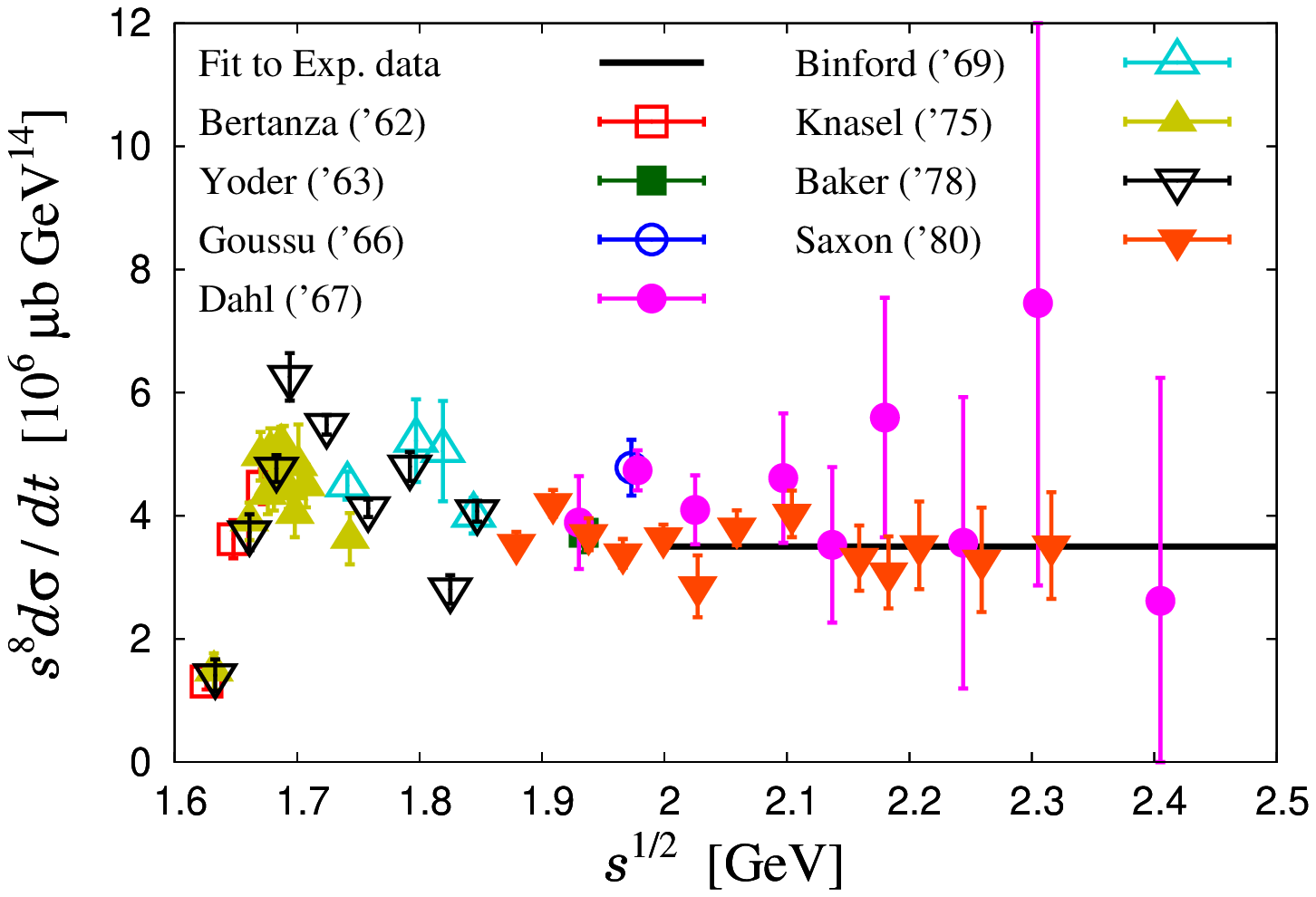}
   \end{center}
\vspace{-0.3cm}
\caption{Cross section of $\pi^- + p \to K^0 + \Lambda$ \cite{kks-2013}.}
\label{fig:lambda-pro}
\end{minipage} 
\end{figure}
%%%%%%%%%%%%%%%%%%%%%%%%%%%%% figure %%%%%%%%%%%%%%%%%%%%%%%%%%%%%

Before discussing the case of $\Lambda \, (1405)$, we first
consider the case of ground-state $\Lambda$, which is expected 
to be an ordinary three-quark system.
There are measurements on the cross section of $\pi^- + p \to K^0 + \Lambda$
\cite{Ronchen:2012eg}.
From the data, we extracted the cross sections at 
$\theta_{c.m.}=90^\circ$, and they are shown in Fig. \ref{fig:lambda-pro-0}.
The data are compared with theoretical estimates calculated 
for the processes $\pi^- + p \to N \, (N^*) \to K^0 + \Lambda$
by including thirteen $s$-channel $N^*$ resonances: 
$N(1535)$, $N(1650)$, $N(1440)$, $N(1710)$, $N(1750)$,
$N(1720)$, $N(1520)$, $N(1675)$, $N(1680)$, $N(1990)$,
$N(2190)$, $N(2250)$, and $N(2220)$
\cite{Ronchen:2012eg}.
There are two curves depending on the parameter choice.
They agree with the data, which indicates that 
the $\Lambda$-production cross sections at low energies
should be described by the formation processes of the $s$-channel 
$N^*$ resonances. 
In order to investigate the high-energy region, the cross section
multiplied by $s^8$ is shown in Fig. \ref{fig:lambda-pro} by
considering the number of total constituents ($n=10$).
Although the experimental data are not very accurate, there is 
a tendency that $s^8 d\sigma/dt$=constant at $\sqrt{s}>2$ GeV.
In fact, fitting the data by the function
$d\sigma / dt = \text{(constant)}  \times s^{2-n}$, 
we obtain the scaling factor \cite{kks-2013}
\begin{align}
n=10.1 \pm 0.6 .
\end{align}
It is consistent with the counting rule, which encourages us
to investigate the scaling behavior of exotic hadron production. 

\begin{figure}[h!]
\begin{minipage}{0.48\textwidth}
   \begin{center}
     \includegraphics[width=6.0cm]{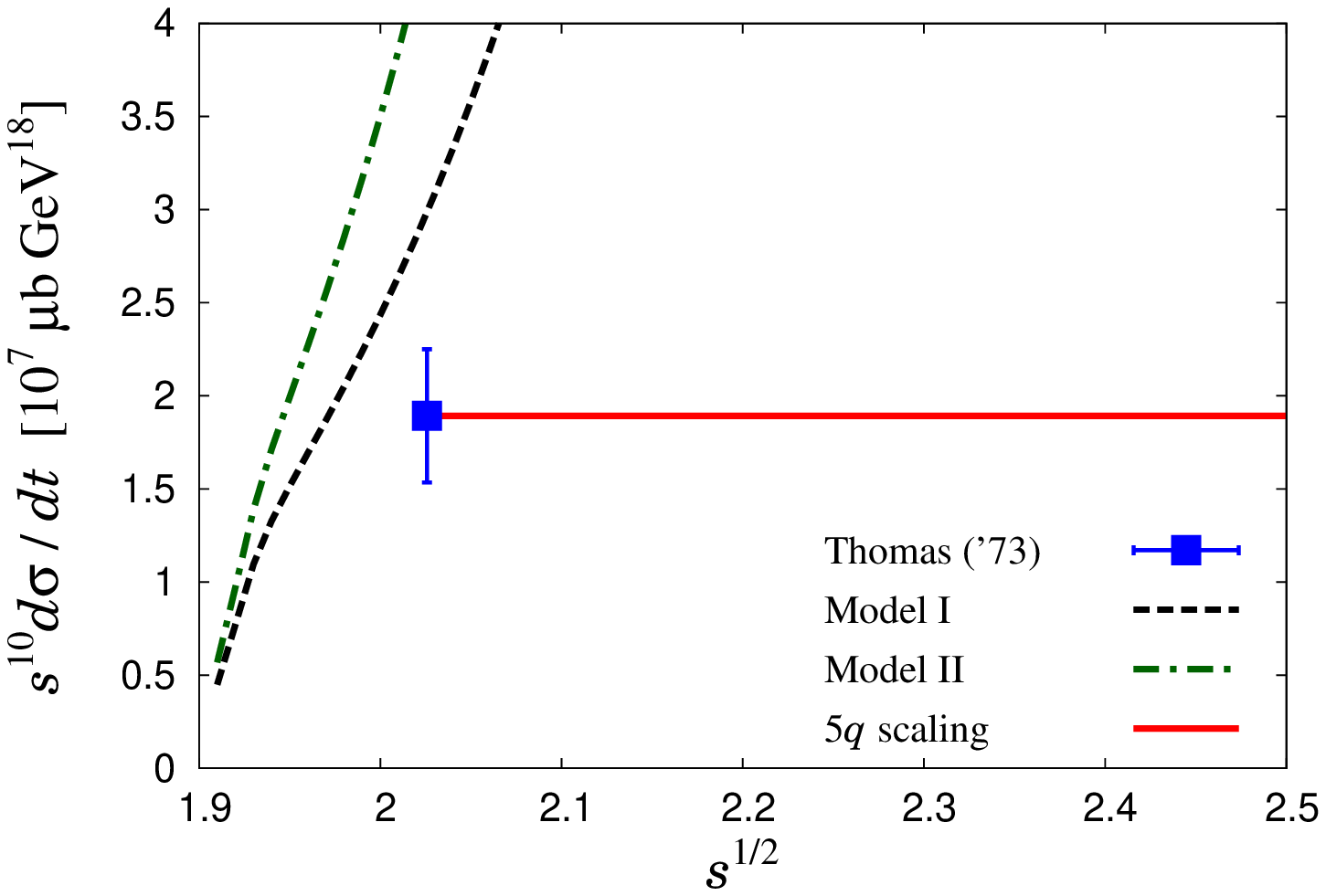}
   \end{center}
\vspace{-0.3cm}
\caption{Cross section of 
         $\pi^- + p \to K^0 + \Lambda \, (1405)$ \cite{kks-2013}.}
\label{fig:lambda-1405-pro}
\end{minipage} 
%%%%%
\hspace{0.5cm}
\begin{minipage}{0.48\textwidth}
   \begin{center}
     \includegraphics[width=6.0cm]{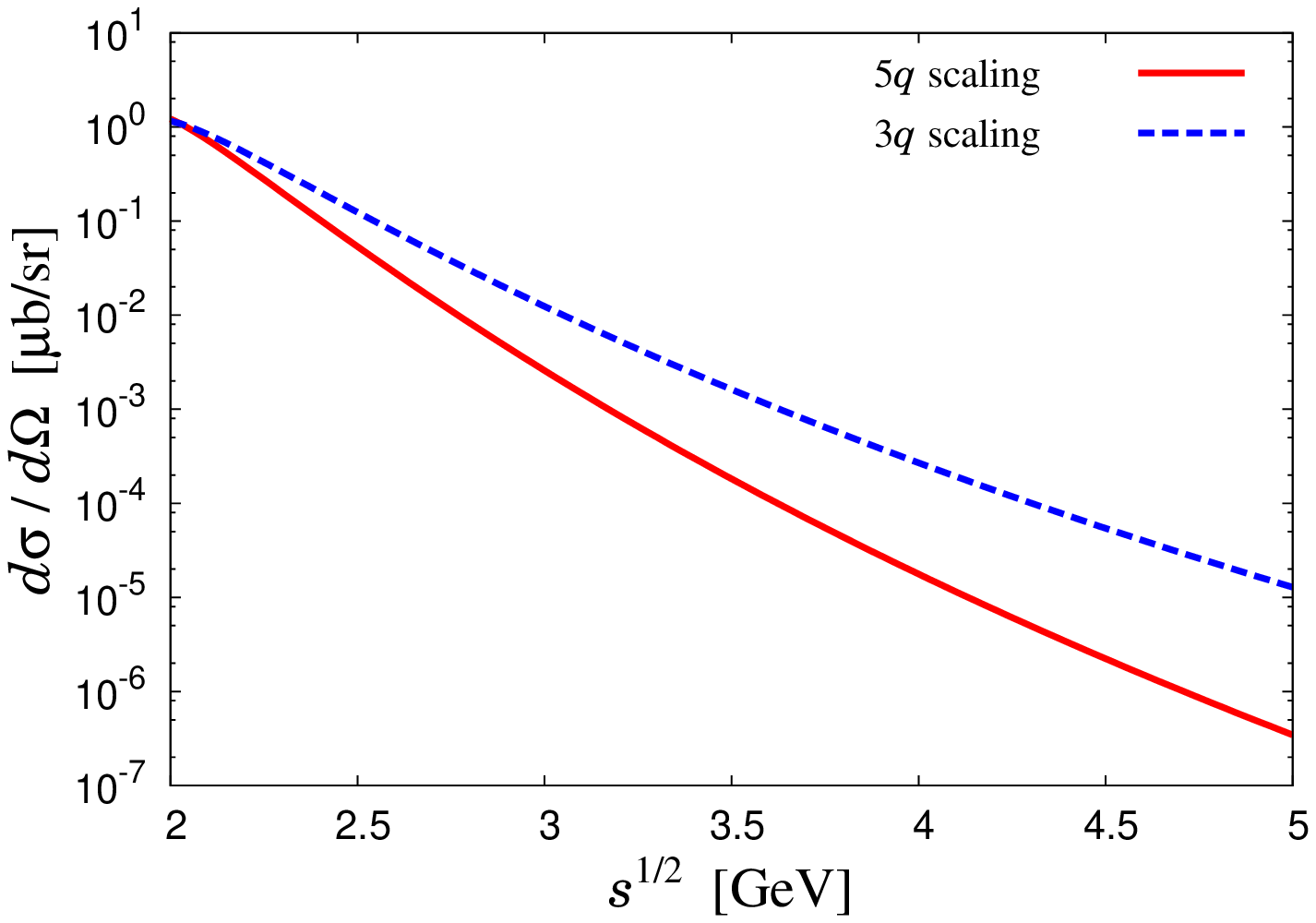}
   \end{center}
\vspace{-0.3cm}
\caption{Scaling of
         $\pi^- + p \to K^0 + \Lambda \, (1405)$ cross section
         at high energies \cite{kks-2013}.}
\label{fig:lambda-1405-scaling}
\end{minipage} 
\end{figure}

Next, we study the cross section of $\pi^- + p \to K^0 + \Lambda (1405)$
for finding the internal structure of $\Lambda \, (1405)$.
However, both experimental and theoretical studies are very limited.
In fact, there is only one experimental measurement 
\cite{lambda-1405-prod-ex} as shown in Fig. \ref{fig:lambda-1405-pro}.
At low energies, the theoretical estimates 
are calculated in the chiral unitary model \cite{lambda-1405-prod-th}, 
which includes meson-exchange
contributions and $s$-channel $N^*(1710)$ formation.
They roughly agree with the data.
For estimating the theoretical cross section at high energies, 
the matrix element of Eq. (\ref{eqn:mab-cd}) should be calculated.
However, it is not easy at this stage because there are a significant
number of processes like Fig. \ref{fig:hard-glun-exchange} should be
calculated, and the distribution amplitudes $\phi_{\pi,p,K^0,\Lambda (1405)}$
are not known. For a rough estimate, we use the experimental data of
Fig. \ref{fig:lambda-1405-pro} and the counting rule. The obtained cross sections
are shown in Fig. \ref{fig:lambda-1405-scaling}.
If the $\Lambda \, (1405)$ is a five-quark state, the total number of
constituents is $n=2+3+2+5=12$. The counting rule indicates 
the scaling $s^{10} d\sigma /dt=$constant, and the cross section
is shown by the solid curve. 
The dashed curve indicates the cross section
if $\Lambda \, (1405)$ is a three-quark baryon.
There is a clear difference between the curves, so that
the internal structure of $\Lambda \, (1405)$ could be
determined by observing the scaling behavior of the exclusive cross section.

We discussed only the $\Lambda \, (1405)$ production; however,
the idea could be applied for other exotic hadron candidates.
They could be experimentally studied at various experimental facilities
including J-PARC, LEP, JLab, and CERN-COMPASS.

\vspace{-0.5cm}
%%%%%%%%%%%%%%%%%%%%%%%%%%%%% figure %%%%%%%%%%%%%%%%%%%%%%%%%%%%%
%%%%%%%%%%%%%%%%%%%%%%%%%%%%% figure %%%%%%%%%%%%%%%%%%%%%%%%%%%%%

\vspace{-0.7cm}
%%%%%%%%%%%%%%%%%%%%%%%%%%%%%%%%%%%%%%%%%%%%%%%%%%%%%%%%%%%%%%%%%%%%%%%%%%%%%%%%
% \appendix
\section*{Acknowledgement}
\vspace{-0.3cm}
This work was partially supported by a Grant-in-Aid for Scientific 
Research on Priority Areas ``Elucidation of New Hadrons with a Variety 
of Flavors (E01: 21105006)" from the ministry of Education, Culture, 
Sports, Science and Technology of Japan.  

%%%%%%%%%%%%%%%%%%%%%%%%%%%%%%%%%%%%%%%%%%%%%%%%%%%%%%%%%%%%%%%%%%%%%%%%%%%%%%%%

%%%%%%%%%%%%%%%%%%%%%%%%%%%%%%%%%%%%%%%%%%%%%%%%%%%%%%%%%%%%%%%%%%%%%%%%%%%%%%%%

\end{document}